\documentclass[doublecol]{epl2}

\usepackage{bm}
\usepackage{epsfig}

\newcommand{\dos}{n}
\newcommand{\ff}{\nu}
\newcommand{\E}{\varepsilon}

\renewcommand{\paragraph}[1]{\textit{#1.---} } 

\newcommand{\diec}{\epsilon_r}

\title{Thermoelectric performance of weakly coupled granular materials}

\author{Andreas~Glatz\inst{1} \and I.~S.~Beloborodov\inst{2}}
\shortauthor{A.~Glatz \etal}
\institute{
  \inst{1} Materials Science Division, Argonne National Laboratory, Argonne, Illinois 60439, USA\\
  \inst{2} Department of Physics and Astronomy, California State University Northridge, Northridge, CA 91330, USA
}

\pacs{73.63.-b}{First pacs description}
\pacs{72.15.Jf}{Second pacs description}
\pacs{73.23.Hk}{Third pacs description}

\date{\today}

\abstract {
We study thermoelectric properties of inhomogeneous nanogranular materials for weak tunneling conductance between the grains, $g_t < 1$. We calculate the thermopower and figure of merit taking into account the shift of the chemical potential and the asymmetry of the density of states in the vicinity of the Fermi surface.
We show that the weak coupling between the grains leads to a high thermopower and low thermal conductivity resulting in relatively high values of the figure of merit on the order of one. We estimate the temperature at which the figure of merit has its maximum value for two- and three-dimensional samples. Our results are applicable for many emerging materials, including artificially self-assembled nanoparticle arrays.
}
\begin{document}

\maketitle

\section{Introduction}
Thermoelectric materials with high efficiency are a major research area in applied physics and materials science for several decades now.
Due to recent advances in nano-fabrication, these materials promise next generation devices for conversion of thermal energy to electrical energy and vice versa.
Especially suited for further improvement in efficiency are {\it inhomogeneous/granular} thermoelectric materials~\cite{Majumdar} in which one can directly control the system parameters.
A measure for the performance or efficiency of a thermoelectric material is the dimensionless {\it figure of merit}, usually denoted as $ZT$, where $T$ is the temperature. It depends on the thermopower or Seebeck coefficient, $S$ and the electric, $\sigma$ and thermal, $\kappa$ conductivities, $ZT = S^2\sigma T/\kappa$.~\cite{Rowe,Mahan}
In Ref.~\cite{venka01} a figure of merit at $300K$ of $2.4$ for a layered nanoscale structure and later in Ref.~\cite{harman02} a $ZT$ of $3.2$ at about $600K$ for a bulk material with nanoscale inclusions were reported.
All these experimental achievements and technological prospects call for a comprehensive theory able to provide a quantitative description of thermoelectric properties of granular materials, which can in future serve as basis for a clever design of devices for a new generation of nano-thermoelectrics. Recently we started this task by studying granular metals~\cite{glatz+prb09a,glatz+prb09b} and here we will answer the question to what extend weak coupling between grains is important in changing $ZT$.

In this Letter we investigate the thermopower $S$ and the figure of merit $ZT$ of granular samples focusing on the case of weak coupling between the grains, $g_t < 1$, see Fig.~1. Each nanoscale cluster is
characterized by (i) the charging energy $E_c=e^2/(\diec a)$, where $e$ is the electron charge, $\diec$ the sample dielectric constant, and $a$ the granule size, and (ii) the mean energy level spacing $\delta$.
The charging energy associated with nanoscale grains can be as large as several thousands Kelvins thus the temperature range $T < E_c$ in which our theory applies, is technologically relevant.

\section{Model and Results}

\begin{figure}
\center\includegraphics[width=0.9\columnwidth]{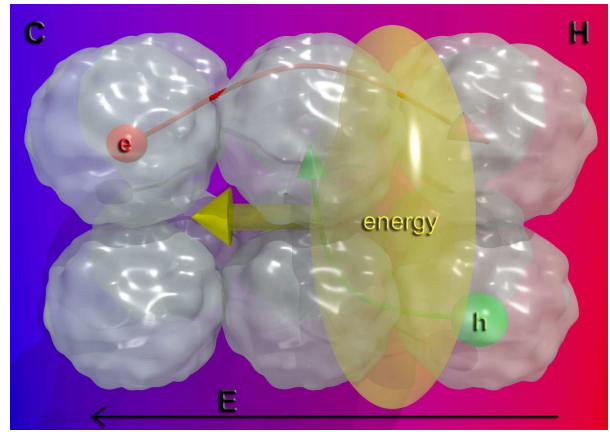}
\caption{(Color online) Sketch of a nanogranular material showing typical electron (e) and hole (h)
trajectories for asymmetric density of states in an electric field ($E$). The energy transport goes from the ''hot'' ($H$) to the ''cold'' ($C$) side.}
\label{fig.model}
\end{figure}

The internal conductance of a  grain is taken much larger than the inter-grain tunneling
conductance, which is the standard condition for granularity.
The tunneling conductance is the main
parameter that controls macroscopic transport properties of the sample~\cite{Beloborodov07}.
In consideration of application to experiments~\cite{venka01,harman02} we restrict ourselves to
the case where the tunneling conductance is smaller than the quantum conductance.

The main results of our work are as follows. (i) We derive the expression for the thermopower $S$ of granular materials in the limit of weak coupling between the grains taking into account a) the shift of the chemical potential $\Delta \mu = a_1 (\ff - 1/2) T_0$ with $\ff$ being the electron filling factor $(|\ff| <1)$, $a_1$ a dimensionless numerical coefficient, and $T_0 = e^2/(\diec \xi_{loc}) \approx E_c$ is the characteristic temperature scale with $\xi_{loc}\simeq a/\ln(E_c^2/T^2g_t)$ being the inelastic localization length~\cite{Beloborodov07} and b) the asymmetry of the density of states (DOS) $\Delta\dos = a_2 (\ff - 1/2)T_0^{-1}$ with $a_2$ being the numerical constant
\begin{equation}
\label{S}
S = - \frac{d\left[\, \Delta\mu +  \Delta\dos T_0 T\right]}{e\, T}=\frac{1/2-\ff}{e}d\left[a_1  \, \frac{T_0}{T} + a_2\right].
\end{equation}
Here $d = 2,3$ is the dimensionality of a sample. We note that the thermopower $S$ of granular materials is finite
{\it only} if the shift of the chemical potential, $\Delta \mu$ and the asymmetry of DOS, $\Delta\dos$ are taken into account. Thus, for filling factor $\ff =1/2$ the thermopower vanishes. 

(ii) We obtain the expression for figure of merit of granular materials in the limit of weak coupling between the grains as follows:
\begin{equation}
\label{ZT}
ZT = \frac{2d^2 g_t e^{-\sqrt{T_0/T}}\left[ \Delta\mu/T +  \Delta\dos T_0\right]^2}{ \gamma_1 g_t^2 (T/E_c)^2 +\gamma_{2,d} (l_{ph}/a)^{2-d} (T/\Theta_D)^{d-1}}  .
\end{equation}
Here $l_{ph}$ is the phonon mean free path, $\Theta_D$ is the Debey temperature, and the parameters $\Delta \mu$ and $\Delta\dos$ are defined above Eq.~(\ref{S}) [$\gamma_1=32\pi^3/15$, $\gamma_{2,2D}=12\zeta(3)/\pi$, and $\gamma_{2,3D}=8\pi^2/15$].
For two and three dimensional samples ($d = 2, 3$) the figure of merit $ZT$ has its maximum value at temperatures $T^*_{d} = T_0/[4(d+1)^2]$~\cite{T0}. We note that $T^*_{2D} > T^*_{3D}$.

\begin{figure}
\includegraphics[width=0.8\columnwidth]{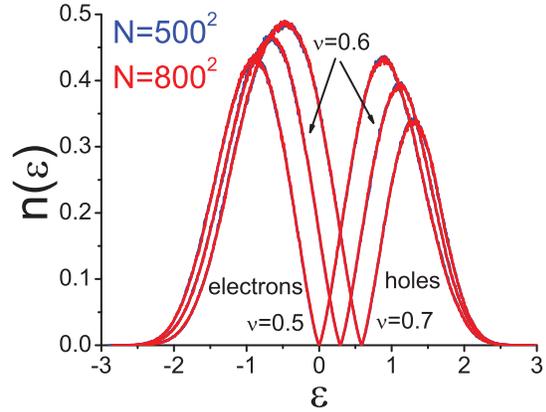}
\caption{(Color online) Result of simulations of the density of state $n(\E)$ vs. energy for different filling factors $\ff$ and different system sizes. The simulations were done for a 2D Coulomb glass model at zero temperature~\cite{shklovskii-book,glatz+prl07,glatz+jstat08}. We chose two large system sizes ($N=500^2$ and $N=800^2$) which show no size dependence for the shift of the chemical potential $\Delta\mu$ and the asymmetry of the density of states $\Delta\dos$. Using simulation data we extract the numerical coefficients $a_1$ and $a_2$ in Eq.~(\ref{S}). We use the Coulomb glass model to simulate the whole weakly coupled granular sample.}
\label{fig.dos}
\end{figure}

Our main results [Eqs.~(\ref{S}) and (\ref{ZT})] are valid for temperatures $T < T_0$ and weak coupling between the grains
$g_t < 1$. Under this conditions, the electric conductivity $\sigma$ and thermal conductivity $\kappa = \kappa_e + \kappa_{ph}$, with $\kappa_e$ and $\kappa_{ph}$ being the electron and phonon contributions to thermal conductivity, are given by expressions~\cite{Beloborodov07,Tripathi,BeloPRB,limit}
\begin{eqnarray}
\label{conductivities}
\sigma  &\simeq& 2e^2 a^{2-d} g_t e^{-\sqrt{T_0/T}}, \\
\kappa_e &\simeq& \gamma_1 a^{2-d} \left[\frac{g_tT}{E_c}\right]^2T , \hspace{0.2cm}
\kappa_{ph} \simeq \gamma_{2,d}l_{ph}^{2-d}  \left[\frac{T}{\Theta_D}\right]^{d-1}T. \nonumber
\end{eqnarray}

We note that the thermopower $S$ of lightly doped semiconductors was investigated in the past~\cite{Fritzsche71,Zvyagin73,zvyagin-pssb73,Kosarev74,Parfenov07}. However, all previous studies were concentrated on the Mott variable range hopping (VRH) regime, with conductivity being $\sigma(T) \sim \exp[-(T_M/T)^{1/4}]$, where $T_M$ is the Mott temperature~\cite{Mottbook}.
In granular materials the Mott VRH regime is hard to observe.
Indeed, the Efros-Shklovskii (ES)
law~\cite{shklovskii-book,ES} may turn into the Mott behavior with an increase of temperature.
This happens when the typical electron energy $\varepsilon$ involved in a hopping
process becomes larger than the width of the Coulomb gap $\Delta_c$, i.e. when it falls into
the flat region of the density of states where Mott behavior is
expected. To estimate the width of the Coulomb gap $\Delta_c$,
one compares the ES expression for the density of states
$\dos(\Delta_c) \sim (\diec/e^2)^d |\Delta_c|^{d-1}$ with the DOS in
the absences of the long-range part of the Coulomb interaction, $n_0$.
Using the condition $\dos(\Delta_c) \sim \dos_0$ we obtain
$\Delta_c = \left( \dos_0 e^{2d}/\diec^d \right) ^{1\over{d-1}  }$. Inserting the value for the bare DOS, $ \dos_0 = 1/E_c \, \xi^d$,
into the last expression we finally obtain $\Delta_c \sim E_c$.
This equation means that there is no flat region in the low temperature DOS.

\section{Thermopower}
Here we show that the thermopower $S$ of granular materials is finite only if the shift of the chemical potential and the asymmetry of DOS are taken into account.
In the following we outline the derivation of Eqs.~(\ref{S}), (\ref{ZT}).

To calculate the thermopower of granular materials in the regime of weak coupling between the grains it is necessary to take into account electrons and holes because the contributions of electrons and holes cancel in the leading order. In general the thermopower is proportional to the average energy transferred by charge carriers and can be written as~\cite{Fritzsche71,Zvyagin73,zvyagin-pssb73,Kosarev74,Parfenov07}
\begin{equation}
\label{1}
S = - \frac{1}{2 e T}\left[ \langle\E - \tilde{\mu}\rangle_e + \langle\E - \tilde{\mu}\rangle_h \right].
\end{equation}
Here the subscripts $e$ and $h$ refer to electrons and holes and $\tilde{\mu} = \mu + \Delta \mu$ is the shifted chemical potential with $\Delta \mu = a_1(\ff - 1/2)T_0$ being the chemical potential shift. The expression in the square brackets of the r.h.s. of Eq.~(\ref{1}) describes the average energy transferred by charge carriers (electron or hole) measured with respect to the shifted chemical potential $\tilde{\mu}$. The average energy in Eq.~(\ref{1}) can be calculated as follows
\begin{equation}
\label{energy}
\langle\E - \tilde{\mu}\rangle_e = \frac{\int \limits_0^{\infty} d\E\, (\E - \tilde{\mu}) \dos(\E - \tilde{\mu}) f(\E - \tilde{\mu}) e^{-\frac{(\E - \tilde{\mu})^2}{2\Delta^2}}}{\int \limits_0^{\infty} d\E\, \dos(\E - \tilde{\mu}) f(\E - \tilde{\mu}) e^{-\frac{(\E - \tilde{\mu})^2}{2\Delta^2}}}.
\end{equation}
Here  $f(\E - \tilde{\mu})$ is the Fermi distribution function, $\Delta = \sqrt{T_0 T}$ is the
typical transfer energy in one hop, and $\dos(\E - \tilde{\mu})$ is the DOS.

Equation~(\ref{energy}) differs in several important ways from previous works~\cite{Fritzsche71,Zvyagin73,zvyagin-pssb73,Kosarev74,Parfenov07}: all considerations in the past were concentrated on the Mott VRH regime with filling factor $\ff = 1/2$. In our consideration the average energy, $\langle\E - \tilde{\mu}\rangle_e+\langle\E - \tilde{\mu}\rangle_h$ is zero for half filling factor. To obtain the finite result in Eq.~(\ref{1}) it is crucial to take into account the asymmetry of the DOS and the shift of the chemical potential.

The DOS, $\dos(\E - \tilde{\mu})$ in Eq.~(\ref{energy}) for $\tilde{\mu} - \Delta_c < \E < \tilde{\mu}$ has the form (see \cite{Zvyagin73,zvyagin-pssb73})
\begin{equation}  \label{dos}
\dos(\E) \propto |\E - \tilde{\mu}|^{d-1}[1 - (\E - \tilde{\mu})\Delta n ]\,,
\end{equation}
and is constant, $n_0$ outside the Coulomb gap region, $\E<\tilde{\mu} - \Delta_c$ and $\E>\tilde{\mu} $, where $\Delta_c \sim E_c$ is the width of the Coulomb gap.
The shift of the chemical potential $\Delta \mu = \tilde{\mu} - \mu$ and the asymmetry of the DOS, $\Delta\dos$, are explicitly defined above Eq.~(\ref{S}).

\begin{figure}
\includegraphics[width=0.8\columnwidth]{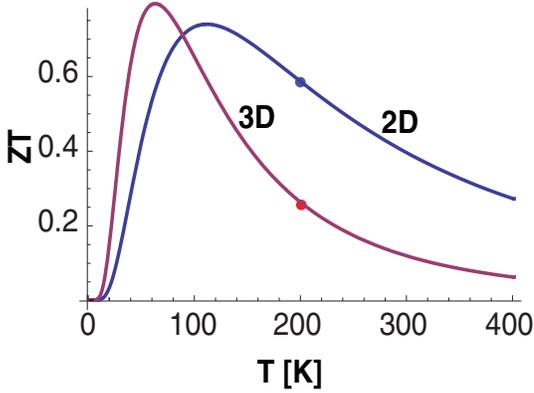}
\caption{(Color online) Plots of figure of merit $ZT$ vs temperature $T$ for two and three dimensional samples. Typical material parameters are given in the text.}
\label{fig.ZT}
\end{figure}

\section{Simulations}

To support our choice for the expression of the DOS $\dos(\E - \tilde{\mu})$ in Eq.~(\ref{dos}) we numerically compute the DOS for a 2D Coulomb glass model~\footnote{For the 3D case we expect similar values for the numerical prefactors $a_{1,2}$, as it was shown in~\cite{glatz+jstat08} for the hard gap in the weakly disordered case.} (see Refs.~\cite{glatz+prl07,glatz+jstat08} for details) at arbitrary filling factor $\ff$ using first principles. The result of the simulations is shown in Fig.~\ref{fig.dos}.
These simulations clearly indicate that for a filling factor $\ff \neq 1/2$, the DOS is asymmetric and the chemical potential is shifted.
By fitting the obtained DOS by expression (\ref{dos}) for different $\ff$ and linear regression of $\Delta\mu(\ff)$ and $\Delta n(\ff)$, we can identify the dimensionless coefficients $a_1\simeq 2.9$ and $a_2\simeq 0.4$.
We note that $a_1\gg a_2$ and therefore the contribution to the thermopower caused by $\Delta\mu$ in
Eq.~(\ref{S}) is dominant.

Now, we can calculate Eq.~(\ref{energy}) and the analog contribution for holes using Eq.~(\ref{dos}) for the DOS together with the numerically found values for $a_{1,2}$. Finally one obtains the result for thermopower $S$ in Eq.~(\ref{S}).
Substituting Eqs.~(\ref{S}) and (\ref{conductivities}) into the expression for the figure of merit $ZT = S^2 \sigma T/(\kappa_e + \kappa_{ph})$  one obtains Eq.~(\ref{ZT}).


\section{Discussions}

In Fig.~\ref{fig.ZT} we plotted the figure of merit $ZT$ for a two- and three-dimensional system, using typical parameters for granular materials: grain size $a \simeq 10 {\rm nm}$ and phonon mean free path $\l_{ph} \simeq a \simeq 1{\rm nm}$. The latter is in general temperature dependent, which can be neglected for the temperatures and grain sizes under consideration.
With $\Theta_D \simeq 450 {\rm K}$, $T_0 \simeq 4000 {\rm K}$, and $g_t \simeq 0.1$ at temperature $T =200 {\rm K}$ using Eq.~(\ref{ZT}) we can simplify $ZT \simeq (2/\gamma_{2,d})(a_1 d)^2 g_t (\ff - 1/2)^2 e^{-\sqrt{T_0/T}}(\Theta_D/T)^{d-1}(T_0/T)^2 $. Using $(\ff - 1/2) = 0.2$ we estimate $(ZT)_{2D} \simeq 0.60$ and $(ZT)_{3D} \simeq 0.25$ (see dots in Fig.~\ref{fig.ZT}).
An important difference between 2D and 3D is that $ZT$ is proportional to $l_{ph}$ in 3D, whereas it is independent in 2D. For the chosen value of $l_{ph}$ both $ZT$ curves are of the same order. However, for larger $l_{ph}$, $ZT$ also becomes larger in $3D$ which is related to the fact that the scattering of phonons in $3D$ is more effective than in $2D$ due to the larger number of grain boundaries. This leads to the suppression of $\kappa_{ph}$ and to higher values of $ZT$.

For hybrid structures of different types of grains this effect might be enhanced and will be subject of a forthcoming work.
Last we mention that in the regime of VRH the contribution due to phonon drag to thermopower is absent~\cite{Zvyagin73,zvyagin-pssb73}.

\section{Conclusions}

In conclusion, we studied thermoelectric properties of inhomogeneous/nanogranular materials at weak tunneling conductance between the grains, $g_t < 1$. We calculated the thermopower $S$ and figure of merit $ZT$ taking into account the shift of the chemical potential and asymmetry of the density of states. We showed that the weak coupling between the grains leads to high thermopower and low thermal conductivity resulting in relatively high values of figure of merit. We calculated the temperature $T^*$ at which the figure of merit has its maximum value.

\acknowledgments
A.~G. was supported by the U.S. Department of Energy Office of Science under the Contract No. DE-AC02-06CH11357.



\begin{thebibliography}{0}
\bibitem{T0} For granular materials the typical temperatur scale $T_0 \sim 4000 {\rm K}$, thus $T^*_{2D} \sim 111 {\rm K}$ and $T^*_{3D} \sim 63 {\rm K}$.
\bibitem{limit} The expression for $\kappa_e$ is valid for temperatures $T > g_t\delta \sim 1 \rm{K}$.
\end{thebibliography}
\end {document}